%% file: style.tex
\documentclass[11pt]{article}

\usepackage{amssymb,amsmath,amsthm}	
\usepackage{a4wide}
\usepackage[utf8x]{luainputenc}               	
\usepackage{graphicx}      										
\usepackage{tikz}          										
\usepackage[authoryear]{natbib}
\usepackage{dsfont}
\usepackage{hyperref}
\usepackage{multirow}
\usepackage{subcaption}
\usepackage{amsmath}
\usepackage{xspace}
\usepackage{cleveref}
\usepackage{wrapfig}
\usepackage{todonotes}
\usepackage{xcolor,colortbl}
\usepackage{color} 

\setcitestyle{round,semicolon,aysep={}}

\newcommand{\klabsatz}{\\[8pt]}          

\newtheorem{example}{Example}[]

\newcommand{\R}{\ensuremath{\mathbb{R}}}

\newcommand{\N}{\ensuremath{\mathbb{N}}}

\begin{document}
	

	\title{ Stylized Facts and Agent-Based Modeling }
	
\author{ Simon Cramer \footnote{WZL, RWTH Aachen University} \footnote{ORCiD IDs: Simon Cramer: 0000-0002-6342-8157, Torsten Trimborn: 0000-0001-5134-7643}, Torsten Trimborn \footnotemark[2] \footnote{IGPM, RWTH Aachen University, Templergraben 55, 52056 Aachen, Germany} \footnote{Corresponding author: trimborn@igpm.rwth-aachen.de}}

\maketitle




\begin{abstract}
The existence of stylized facts in financial data has been documented in many studies.
In the past decade the modeling of financial markets by agent-based computational economic market models has become a frequently used modeling approach.
The main purpose of these models is to replicate stylized facts and to identify sufficient conditions for their creations. 
In this paper we introduce the most prominent examples of stylized facts and especially present stylized facts of financial data.
Furthermore, we given an introduction to agent-based modeling.
Here, we not only provide an overview of this topic but introduce the idea of universal building blocks for agent-based economic market models. 
\end{abstract}

\include{introduction}

\include{method}
\include{conclusion}
\section*{Acknowledgement}
 S. Cramer and T. Trimborn were funded by the Deutsche Forschungsgemeinschaft (DFG, German Research Foundation) under Germany's Excellence Strategy – EXC-2023 Internet of Production – 390621612.\\
T. Trimborn gratefully acknowledges support by the Hans-Böckler-Stiftung and the RWTH Aachen University Start-Up grant. 
T. Trimborn acknowledges the support by the ERS Prep Fund - Simulation and Data Science. 
The work was partially funded by the Excellence Initiative of the German federal and state governments.
\appendix


	\bibliography{Quellen/SABCEMM.bib}
	\bibliographystyle{plainnat}

\end{document}

%% file: introduction.tex
\section{Introduction}
The purpose of this review paper is to give an introduction to stylized facts and agent-based economic market models.
We review the most important stylized facts and compute several statistical measures on financial data.
In terms of agent-based models, we focus on the modeling of agent-based economic market models and the relevant concepts.
\\ \\
This paper is structured as follows:
In the next section, we provide an introduction to stylized facts and present the empirical results for DAX, S\&P and Dow Jones data.
In section 3 we introduce the idea of agent-based modeling and aim for a short overview of this vivid research area.
In addition, we present universal building blocks for agent-based economic market (ABCEM) models.
This shall be seen as a modeling framework and assists in gaining a universal perspective on ABCEM models.
Finally, we finish this paper with a short conclusion.

\section{Stylized Facts}
\begin{figure}[h!]
\begin{center}
\includegraphics[width=0.48\textwidth]{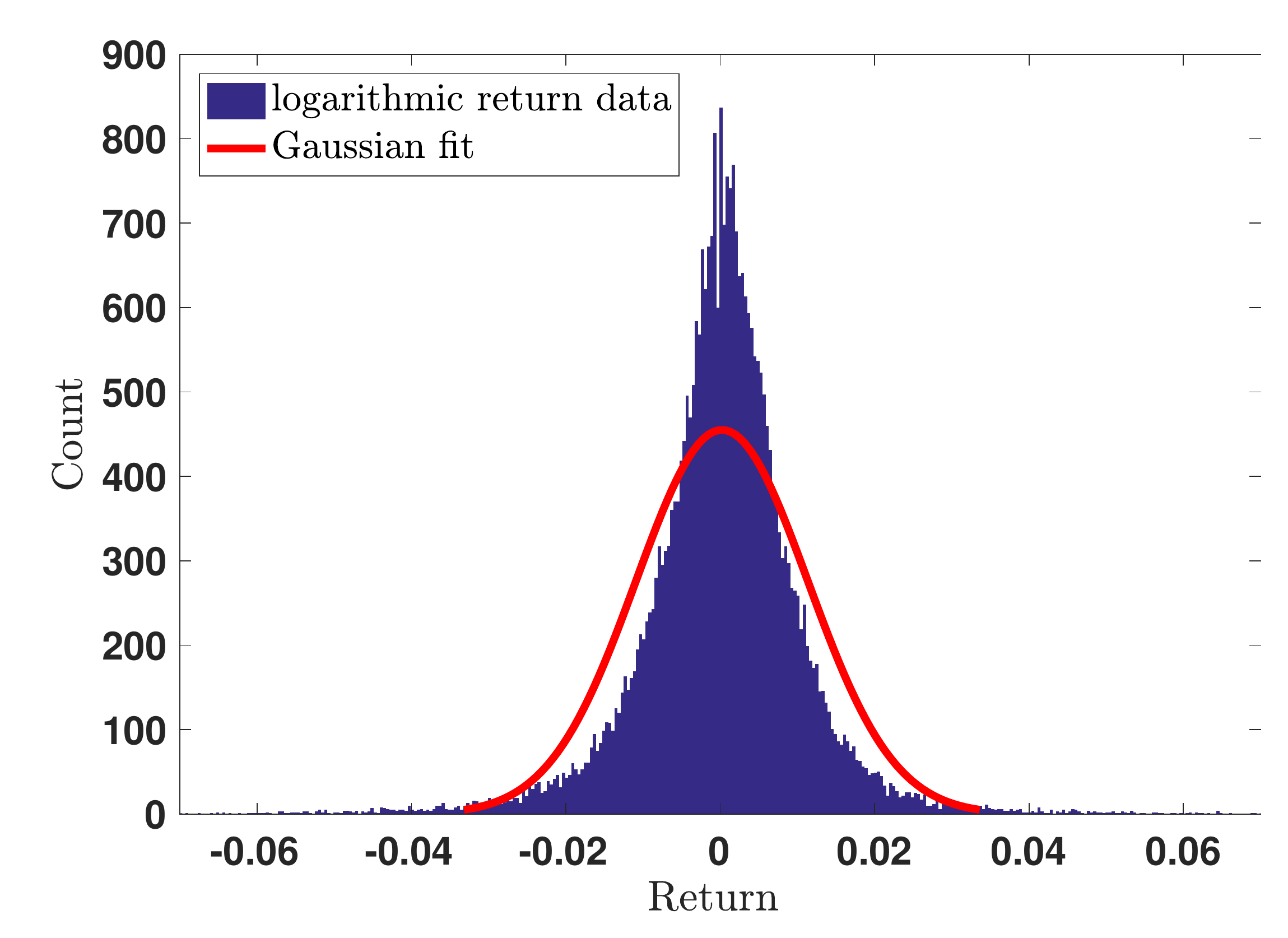}
\hfill
\includegraphics[width=0.48\textwidth]{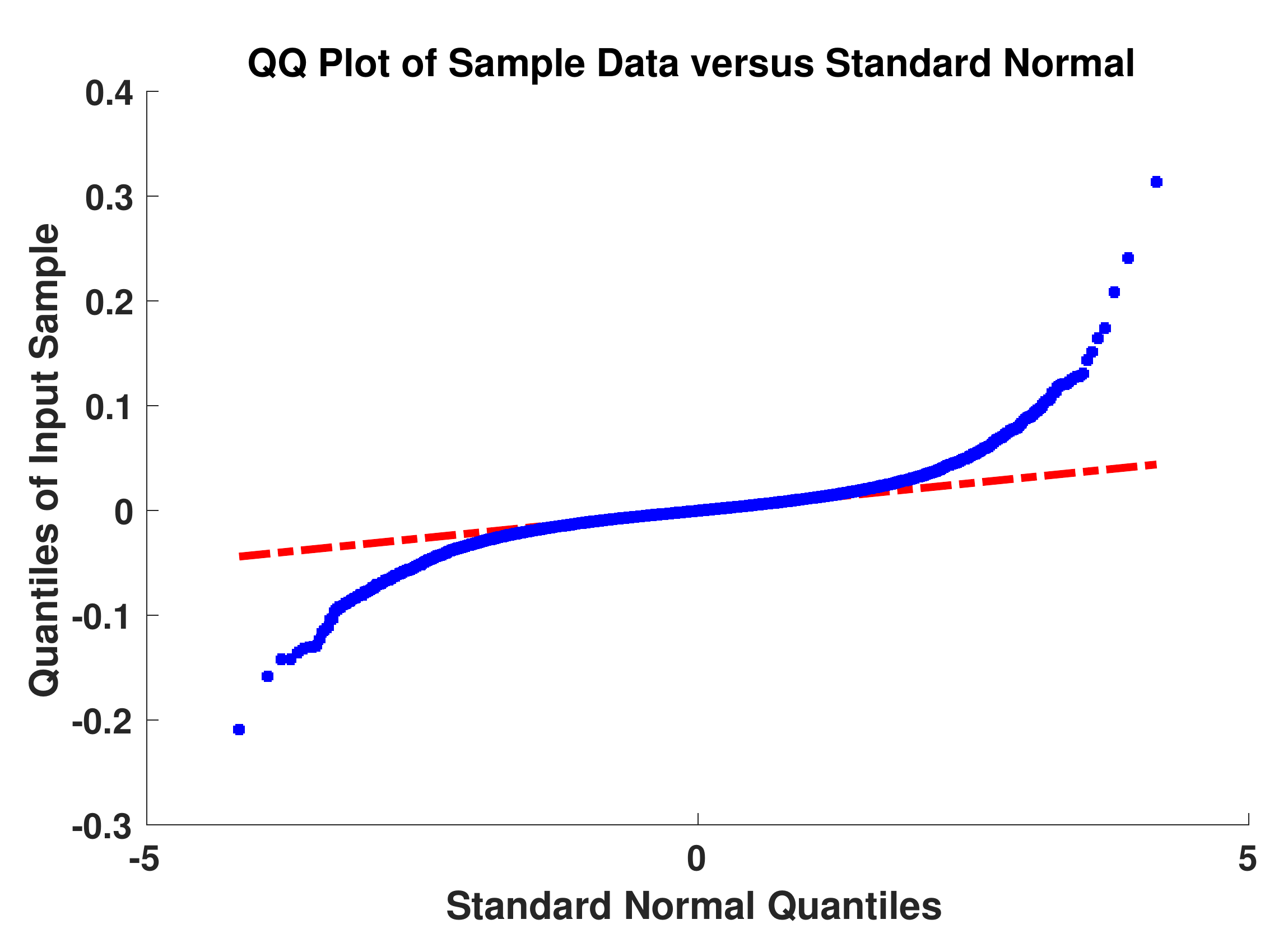}
\caption{Histogram of daily Dow Jones data with a Gaussian fit (left hand side) and a quantile-quantile plot of the logarithmic return data. Dow Jones data (May 27, 1896-November 14, 2018). Source: \texttt{https://stooq.com} (accessed on November 15, 2018). }\label{hist}
\end{center}
\end{figure}

The empirical observation of stylized facts in financial data dates back more than 100 years.
The observation of inequality in income may be seen as the first stylized fact documented by Vilfredo Pareto in 1897 \citep{pareto1897cours}.
Stylized facts are commonly accepted as persistent empirical patterns in financial data.
Furthermore, they are universal in that sense that they can be  observed on different markets and even on different time scales all over the world.\\ 
The first true empirical observations of stylized facts has probably been done by Fama and Mandelbrot in the 1960s \citep{mandelbrot1997variation, brada1966letter, FamaPhd}.
They have shown that the stock return distribution is not well fitted by a Gaussian distribution and obtained the well-known fat-tail characteristic of stock return data.
This stylized fact can be quantified by the inverse cumulative distribution function of logarithmic stock returns $F_c(r),\ r\in \R$,  where we denote the corresponding random variable $R\in\R$:
$$
F_c(r):= \int\limits_r^{\infty} \Phi(\tilde{r})\ d\tilde{r} \sim \frac{1}{r^{\mu}},\quad \mu>0.
$$
Here, $\Phi$ denotes the distribution function of the logarithmic stock return distribution and $\mu$ the Pareto exponent.
The question how to quantify the deviations from Gaussianity is very crucial. 
One frequently used measure  of the fatness of the tail and the peak at the mean of a distribution is the excess kurtosis, given as the normalized fourth moment of stock returns $R$  minus a correction term, defined by
\begin{align*}
&\kappa:= \frac{E[(R-\bar{R})^2]}{\sigma^4}-3,\\
& \bar{R}:= E[R],\\
&\sigma^2:= E[(R-\bar{R})^2]. 
\end{align*}
The correction term is needed to obtain Gaussian behavior for $\kappa=0$ called mesokurtic.
The stock return distribution exhibits leptokurtic behavior which corresponds to $\kappa>0$. 
Thus, the stock return distribution has a higher peak around the mean value and a heavier tail than the Gaussian distribution as Figure \ref{hist} reveals. 
The tail exponent of the inverse cumulative distribution function of logarithmic stock returns can be estimated by the Hill estimator (see appendix definition \ref{def:hill_estimator}).
Examples of the excess kurtosis and the Hill estimator of real stock price data is given in Tables \ref{tab:market_data}-\ref{tab:market_data_absret_20y}.

\begin{figure}[h!]
\begin{center}
\includegraphics[width=.7\textwidth]{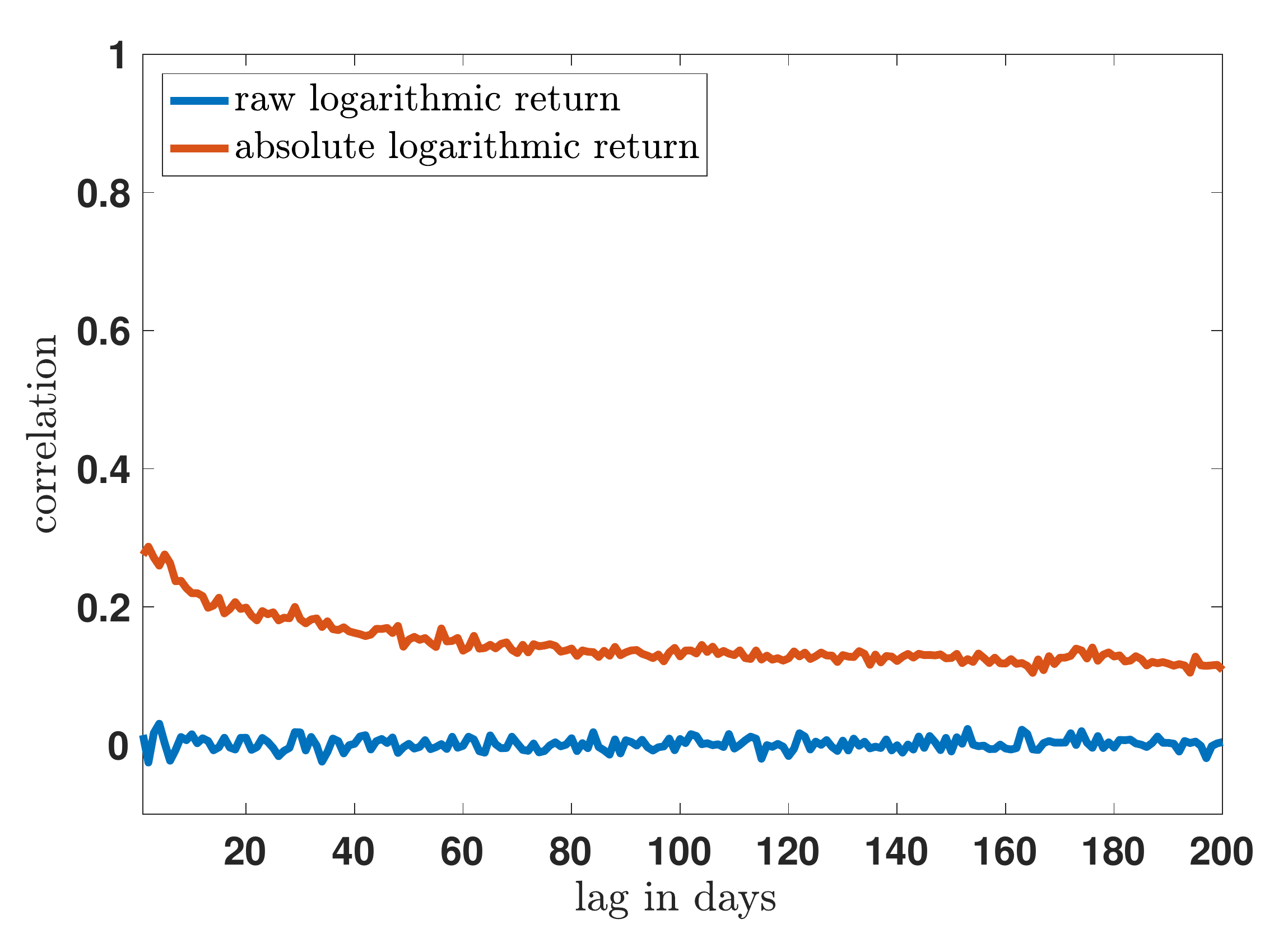}
\caption{Auto-correlation of raw and absolute returns of daily Dow Jones data (May 27, 1896-November 14, 2018).  Source: \texttt{https://stooq.com} (accessed on November 15th 2018). }\label{AutoInt}
\end{center}
\end{figure}

A further possibility for visualization of the fat-tail property of stock returns is to make use of quantile-quantile plots (qq-plots).
The qq-plot in figure \ref{hist} plots the data against a Gaussian distributions and the deviations from the straight line clearly indicates the fat-tail behavior.
\\ \\
A further example of a stylized fact in stock prices is volatility clustering, obtained again by Mandelbrot \citep{mandelbrot1997variation}.
This stylized fact can be quantified with the help of the auto-correlation function.
Since the stock return distribution is assumed to be a stationary stochastic process one can calculate the correlation between stock returns at different points in time.
The auto-correlation for the stationary stochastic process $R(t),\ t>0$ is given by:
\begin{align*}
C(l):= Corr(R(t+l),R(t))=\frac{Cov(R(t+l), R(t))}{E[(R(t)-\bar{R})^2]}= \frac{E[(R(t+l)-\bar{R})\ (R(t)-\bar{R})]}{E[(R(t)-\bar{R})^2]},\ l>0
\end{align*}
The correlation $Corr$ is given by the normalized covariance $Cov$ of two random variables.
The auto-correlation function $C(l)\in[-1,1]$ depends on the time shift called lag $l>0$ of the stochastic process.
Empirical data reveals that raw returns are not auto-correlated but in fact absolute or quadratic raw returns possess significant auto-correlation.
Interestingly, in the case of absolute or squared returns the auto-correlation function exhibits an algebraic decay similar to the stock return distribution.
$$
 Corr(|R(t+l)|,|R(t)|)\sim \frac{1}{l^{\beta}},\  \beta>0. 
$$
The Figure \ref{AutoInt}  depicts the empirical auto-correlation function of raw and absolute daily returns of DAX data.
We clearly obtain no auto-correlation for raw returns and a slowly decreasing auto-correlation with respect to the time lag for absolute returns.
\\ \\

Besides the previously presented stylized facts there are at least thirty stylized facts documented \citep{chen2012agent, lux2008stochastic}.
For a detailed discussion on stylized facts we refer to  \citep{cont2001empirical, ehrentreich2007agent, campbell1997econometrics, pagan1996econometrics, lux2008stochastic}.
\\ \\
Although stylized facts are ``almost universally accepted among economists and physicists" \citep{maldarella2012kinetic},  the origins of stylized facts remain widely undiscovered \citep{pagan1996econometrics, cowan2002heterogenous, maldarella2012kinetic}.
For example many stylized facts cannot be explained by the famous Efficient Market Hypothesis by Eugene Fama \citep{fama1965behavior} which has been the dominant paradigm in macroeconomics for many years.
Furthermore, several studies indicate that stylized facts play a crucial role in the creation of financial crashes \citep{farmer2009economy, lebaron2006agent}.

\subsection{Empirical Data}
The empirical results for the excess kurtosis, Hill estimator and auto-correlations for the daily logarithmic stock returns and absolute logarithmic stock returns of the  indices DAX, Dow Jones and S\&P have been computed.
We review two data sources, namely \texttt{Yahoo} and \texttt{Stooq}.
The Hill estimator is calculated on the upper 5\% of the stock return data.
The auto-correlation is evaluated for  different time lags $(10,20,50, 100)$.
Furthermore, we consider different time horizons of our indices as presented in the Tables \ref{tab:market_data}-\ref{tab:market_data_absret_20y}.
\\ \\
The data reveals that the excess kurtosis is very sensitive with respect to different indices.
Thus, we obtain differences in the excess kurtosis of  DAX and S\&P data with a factor of more than four (see Table \ref{tab:market_data} ).
In addition, these tables reveal that the time horizon has an impact on all other statistical quantities as well.
For example one may compare Table \ref{tab:market_data} with Table \ref{tab:market_data_logret_10y}  or Table  \ref{tab:market_data_absret_10y} with Table \ref{tab:market_data_absret_20y}. 
Finally, we aim to discuss the quality of the data.
For all time horizons, where data of both sources were available,  we presented them.
The results indicate that the quality of the data appears sufficient.
Differences are only obtained in an order of magnitude $10^{-2}$.
Nevertheless, we have marked the largest differences (see Tables \ref{tab:market_data_logret_10y}, \ref{tab:market_data_logret_20y}).

\begin{table}
    \centering
    \begin{tabular}{|c|c|c|c|c|c|}
        \hline 
        & DAX 1959-2018 &  DAX 1980-2018& DJ 1896-2018 & SP 1789-2018 & SP 1980-2018 \\ 
        \hline 
        Skew & -0.15398 & -0.29760 & -0.53018 & 0.04980 & -1.15998 \\ 
        \hline 
        Excess Kurtosis& 7.48509 & 6.41534 & 19.94971 & 32.49845 & 26.80929 \\ 
        \hline 
        Hill 5\% & 2.86114 & 2.95151 & 2.61625 & 2.38089 & 2.87303 \\ 
        \hline
        AutoCorr 10 & 0.00097 & -0.00215 & 0.01552 & 0.00801 & 0.01431\\ 
        \hline 
        AutoCorr 20 & 0.00993 & 0.01534 & 0.01043 & 0.00537 & 0.01499 \\ 
        \hline
        AutoCorr 50 & -0.00689 & -0.00663 & 0.00169 & -0.00973 & -0.02552 \\ 
        \hline
        AutoCorr 100 & -0.00382 & -0.00081 & 0.00866 & 0.00432 & 0.01582 \\ 
        \hline 
    \end{tabular}
\caption{Market Data: \textbf{Log returns} from Stooqs daily open prices}
\label{tab:market_data}
\end{table}

\begin{table}
    \centering
    \begin{tabular}{|c||c|c||c|c||c|c||}
        \hline 
        & \multicolumn{2}{c||}{DAX} &  \multicolumn{2}{c||}{Dow Jones} & \multicolumn{2}{c||}{S\&P 500} \\ 
        \hline
        & Stooqs & Yahoo & Stooqs & Yahoo & Stooqs & Yahoo \\
        \hline 
        Hill 5\% & 3.27487 & 3.27484 & 2.36088 & 2.36088 & 2.25837 & 2.25842 \\
        \hline 
        Excess Kurtosis & 5.85989 & 5.86956 & 10.9572 & 10.9333 & 11.1009 & 11.0984 \\  
        \hline 
        Skew & \cellcolor{yellow!50}0.00476301 & \cellcolor{yellow!50}-0.00979905 & -0.0866539 & -0.0858895 & -0.340703 & -0.340731 \\ 
        \hline
        AutoCorr 10 & \cellcolor{yellow!50}0.0045713 & \cellcolor{yellow!50}0.0143741 & 0.0275151 & 0.0276896 & 0.0351904 & 0.0356558 \\ 
        \hline 
        AutoCorr 20 & \cellcolor{yellow!50}0.0133577 & \cellcolor{yellow!50}0.00854401 & 0.0557108 & 0.0547723 & 0.0614776 & 0.0606093 \\ 
        \hline
        AutoCorr 50 & \cellcolor{yellow!50}-0.0677308 & \cellcolor{yellow!50}-0.0366783 & -0.0473323 & -0.0463268 & -0.0573844 & -0.0554321 \\ 
        \hline
        AutoCorr 100 & \cellcolor{yellow!50} 0.0224124 & \cellcolor{yellow!50}0.0131171 & 0.0128191 & 0.0126586 & 0.0227003 & 0.0222299 \\ 
        \hline 
    \end{tabular}
    \caption{Market Data: \textbf{Log returns} of open prices for 10 years (1/1/2008 - 1/1/2018) with about 2538 trading days. Large Deviations are highlighted.}
    \label{tab:market_data_logret_10y}
\end{table}

\begin{table}
    \centering
    \begin{tabular}{|c||c|c||c|c||c|c||}
        \hline 
        & \multicolumn{2}{c||}{DAX} &  \multicolumn{2}{c||}{Dow Jones} & \multicolumn{2}{c||}{S\&P 500} \\ 
        \hline
        & Stooqs & Yahoo & Stooqs & Yahoo & Stooqs & Yahoo \\
        \hline 
        Hill 5\% & 3.12956 & 3.08732 & 2.77032 & 2.7728 & 2.506 & 2.50589 \\
        \hline 
        Excess Kurtosis & 14.1932 & 14.1195 & 22.0448 & 22.0289 & 20.3362 & 20.3454 \\  
        \hline 
        Skew & 2.78073 & 2.78152 & 3.54775 & 3.54489 & 3.5181 & 3.5183 \\ 
        \hline
        AutoCorr 10 & 0.206428 & 0.205136 & 0.322451 & 0.321833 & 0.334737 & 0.33453 \\ 
        \hline 
        AutoCorr 20 & 0.16254 & 0.172491 & 0.280044 & 0.2803 & 0.286463 & 0.286256 \\ 
        \hline
        AutoCorr 50 & 0.0876169 & 0.0711959 & 0.173887 & 0.174129 & 0.180332 & 0.180348 \\ 
        \hline
        AutoCorr 100 &  0.0771358 & 0.0793103 & 0.144337 & 0.143757 & 0.142684 & 0.142319 \\ 
        \hline 
    \end{tabular}
    \caption{Market Data: \textbf{Abs log returns} of open prices for 10 years (1/1/2008 - 1/1/2018) with about 2538 trading days.}
    \label{tab:market_data_absret_10y}
\end{table}

\begin{table}
    \centering
    \begin{tabular}{|c||c|c||c|c||c|c||}
        \hline 
        & \multicolumn{2}{c||}{DAX} &  \multicolumn{2}{c||}{Dow Jones} & \multicolumn{2}{c||}{S\&P 500} \\ 
        \hline
        & Stooqs & Yahoo & Stooqs & Yahoo & Stooqs & Yahoo \\
        \hline 
        Hill 5\% & 3.0432 & 3.02615 & 2.72374 & 2.7234 & 2.56005 & 2.56067 \\
        \hline 
        Excess Kurtosis & 4.11224 & 4.14143 & 7.96191 & 7.95903 & 8.08028 & 8.09152 \\  
        \hline 
        Skew & -0.0936016 & -0.0868356 & -0.115667 & -0.115951 & -0.217571 & -0.218462 \\ 
        \hline
        AutoCorr 10 & \cellcolor{yellow!50}-0.0126055 & \cellcolor{yellow!50}-0.00218874 & 0.0221733 & 0.0222847 & 0.0236101 & 0.0237844 \\ 
        \hline 
        AutoCorr 20 & 0.0215986 & 0.0208578 & 0.0149777 & 0.0151891 & 0.0177763 & 0.0177882 \\ 
        \hline
        AutoCorr 50 & \cellcolor{yellow!50}-0.0228001 & \cellcolor{yellow!50}-0.00978986 & -0.0311535 & -0.0310373 & -0.0366033 & -0.0360126 \\ 
        \hline
        AutoCorr 100 &\cellcolor{yellow!50} 0.0104454 &\cellcolor{yellow!50} 0.00583078 & 0.020815 & 0.0209226 & 0.0235104 & 0.0232849 \\ 
        \hline 
    \end{tabular}
    \caption{Market Data: \textbf{Log returns} of open prices for 20 years (1/1/1998 - 1/1/2018) with about 5079 trading days. Large Deviations are highlighted.}
    \label{tab:market_data_logret_20y}
\end{table}

\begin{table}
    \centering
    \begin{tabular}{|c||c|c||c|c||c|c||}
        \hline 
        & \multicolumn{2}{c||}{DAX} &  \multicolumn{2}{c||}{Dow Jones} & \multicolumn{2}{c||}{S\&P 500} \\ 
        \hline
        & Stooqs & Yahoo & Stooqs & Yahoo & Stooqs & Yahoo \\
        \hline 
        Hill 5\% & 3.30068 & 3.24861 & 2.95592 & 2.95606 & 2.95044 & 2.95047 \\
        \hline 
        Excess Kurtosis & 9.64775 & 9.68247 & 18.3816 & 18.3859 & 17.5795 & 17.6036 \\  
        \hline 
        Skew & 2.36843 & 2.37762 & 3.10218 & 3.10216 & 3.08068 & 3.08304 \\ 
        \hline
        AutoCorr 10 & 0.245037 & 0.248828 & 0.272268 & 0.27209 & 0.288187 & 0.288096 \\ 
        \hline 
        AutoCorr 20 & 0.201899 & 0.201428 & 0.227346 & 0.227286 & 0.236044 & 0.235682 \\ 
        \hline
        AutoCorr 50 & 0.153015 & 0.145099 & 0.155416 & 0.155355 & 0.167528 & 0.167349 \\ 
        \hline
        AutoCorr 100 & 0.118164 & 0.118089 & 0.117543 & 0.117262 & 0.121554 & 0.12127 \\ 
        \hline 
    \end{tabular}
    \caption{Market Data: \textbf{Abs log returns} of open prices for 20 years (1/1/1998 - 1/1/2018) with about 5079 trading days.}
    \label{tab:market_data_absret_20y}
\end{table}

\clearpage

\section{Agent-Based Models}
One possible approach to gain insights into the creation of stylized facts are computational agent-based models which are part of the research field econophysics.
This approach borrows tools from statistical mechanics, such as Monte-Carlo simulations, and are often inspired by physical theories or models such as kinetic theory or the famous Ising model. 
Agent-based modeling has become very popular modeling tool over the last decade \citep{farmer2009economy, hommes2006heterogeneous, tesfatsion2002agent}.
Such model can be applied to nearly all fields of economics, examples include wealth formation, firm size, stock market, policy design, innovation change or auction markets.  
The starting point of the first modern multi-agent model \citep{samanidou2007agent} is probably the market crash of 20\% at the US stock market in 1987.
This extreme anomaly known as \textit{Black Monday}, which economists failed to provide an explanation for, encouraged the economists Kim and Markowitz to design an agent-based model \citep{kim1989investment}.
They tried to discover connections between agents who follow a portfolio insurance strategy and the volatility of the market with the help of Monte-Carlo simulations.
\\ \\
These modern financial market models of interacting heterogeneous agents share many similarities with interacting particle systems from physics \citep{sornette2014physics, zschischang2001some, lux2008applications}. 
These models usually consider bounded rational agents in the sense of Simon \citep{simon1955behavioral} and are influenced by behavioral finance \citep{lebaron2006agent, farmer2009economy, hommes2006heterogeneous, chen2012agent}.
Many agent-based financial market models are able to replicate the most prominent stylized facts of financial markets such as fat-tails of asset returns or volatility clustering.
Thus, these computational agent-based models are able to shed light on the origins of stylized facts.
More precisely they are able to find sufficient conditions for the agent or market design in order to obtain stylized facts.
\\ \\
As many studies indicate behavioral aspects of financial agents may be one reason for the creation of stylized facts \citep{cross2005threshold, lux2008stochastic, chen2012agent}.
New theories have been developed such as the interacting agent hypothesis \citep{lux1999scaling, ehrentreich2007agent} or the heterogeneous market hypothesis proposed by Hommes \citep{hommes2001financial, ehrentreich2007agent} as alternatives to the efficient market hypothesis.
For a comprehensive introduction to agent-based models we refer to \citep{ehrentreich2007agent, chen2012agent, janssen2006empirically, cont2007volatility, lebaron2000agent, lebaron2006agent, samanidou2007agent, hommes2006heterogeneous, iori2012agent, sornette2014physics}.
\\ \\
The great advantage of agent-based models compared to traditional models is the possibility to design complex agents and study the interaction of those by computer simulations.
As the name reveals, the modeling of the agent is the key aspect.
Modeling financial agents has a long tradition in economics and dates back the work of Smith \citep{smith1937wealth}.
Recent developments in the field of behavioral finance had a significant influence on agent-based models.
In the next paragraph, we will provide a short overview of agent modeling and introduce the concept of bounded rational agents which is used in most agent-based models. 
 
\paragraph{Modeling Agents}
The question of modeling financial agents is actually concerned with the modeling of a decision-making process. 
Thus, in the case of an financial market, agents are faced with the decision to buy, hold or sell a stock (good) or to be flat in the market.
The theory of choice is known in economics as decision theory or utility theory.
Besides early contributions of Bentham, Gossen and Depuit, modern utility theory has been developed by Walras and Menger \citep{stigler1950development}.
These early studies focused on proper utility measures and utility maximization.
Notable contributions on the mathematization of utility models have been published by Edgeworth \citep{edgeworth1881mathematical}. \klabsatz 
In the context of financial market models respectively agent-based models, we are especially interested in the theory of expected utility also known as expected utility hypothesis.
This theory deals with the modeling of the decision process of persons under uncertain outcomes.
The first example for the problem of choice under uncertainties was given by Bernoulli in 1713 with the famous St. Petersburg paradox.
In the 1930s and 1940s, the expected utility hypothesis has been put on a solid mathematical foundation by the mathematician von Neumann and economist Morgenstern.
In 1944, they published the famous von Neumann-Morgenstern utility theorem \citep{von2007theory} which precisely defines when a decision maker is rational, i.e. if a utility function and the corresponding maximum exist.\klabsatz 
The model of rational expectation of financial agents has been rigorously defined by Muth in 1961 \citep{muth1961rational} and is known as the rational expectation hypothesis.
It says that the agents' expectation (e.g. of the future stock price) is equal to the true expected value of the economic asset.
Thus, the agents' expectations may deviate from the correct value but is true on the average.
This theory became the dominant macroeconomic approach after Lucas used it in his famous Lucas critique in 1976 \citep{lucas1976econometric, ehrentreich2008agent}.
Furthermore, the rational expectation hypothesis is the foundation of the famous efficient market hypothesis by E. Fama \citep{malkiel1970efficient}. \klabsatz
As discussed earlier, market models of rational agents are not able to explain and reproduce stylized facts.
For that reason agent-based models do not follow the rational expectation hypothesis, they follow the ansatz of boundedly rational agents.
Thus, also behavioral aspects are considered in the decision making of the agents.

\paragraph{Bounded Rational Agents}
The concept of bounded rational agents has been introduced by Simon \citep{simon1955behavioral, simon1957models}. 
Like the theory of expected utility, this is a model of the agents' choice.
Bounded rational agents do not only act rational but partly irrational.
Mathematically, they do not solve an optimization problem, but rather look for a satisfactory solution which is near the optimum.
This can be supported by the fact that the computational resources, respectively the time to solve the optimization problem are limited in real world application \citep{ehrentreich2008agent}.
Furthermore, it is well known that fund managers often prefer to apply heuristics than to solve a highly complex optimization problem.
One extension of this model has been derived by Rubinstein  \citep{rubinstein1998modeling}.
The concept of bounded rational agents is heavily influenced and supported by behavioral finance.
Thus, the deviations of financial agents to the optimal solution can be accounted for behavioral biases of agents.
Probably the most famous theory in behavioral finance is the so called prospect theory which has been established by Kahnemann and Tversky in their seminal paper \textit{Prospect Theory: An Analysis of Decision under Risk} in 1979.
Prospect theory deals with the decision making of agents under uncertain outcomes.
It attempts to approximate real-life heuristics of decision makers which are influenced by psychological effects. 
In some sense, this theory can be seen as an extension of the expected utility theory.

%% file: method.tex
\subsection{Abstract Agent-Based Economic Market Model}
\label{sabcemm}
\begin{figure}[h!]
\begin{center}
\includegraphics[width=0.5\textwidth]{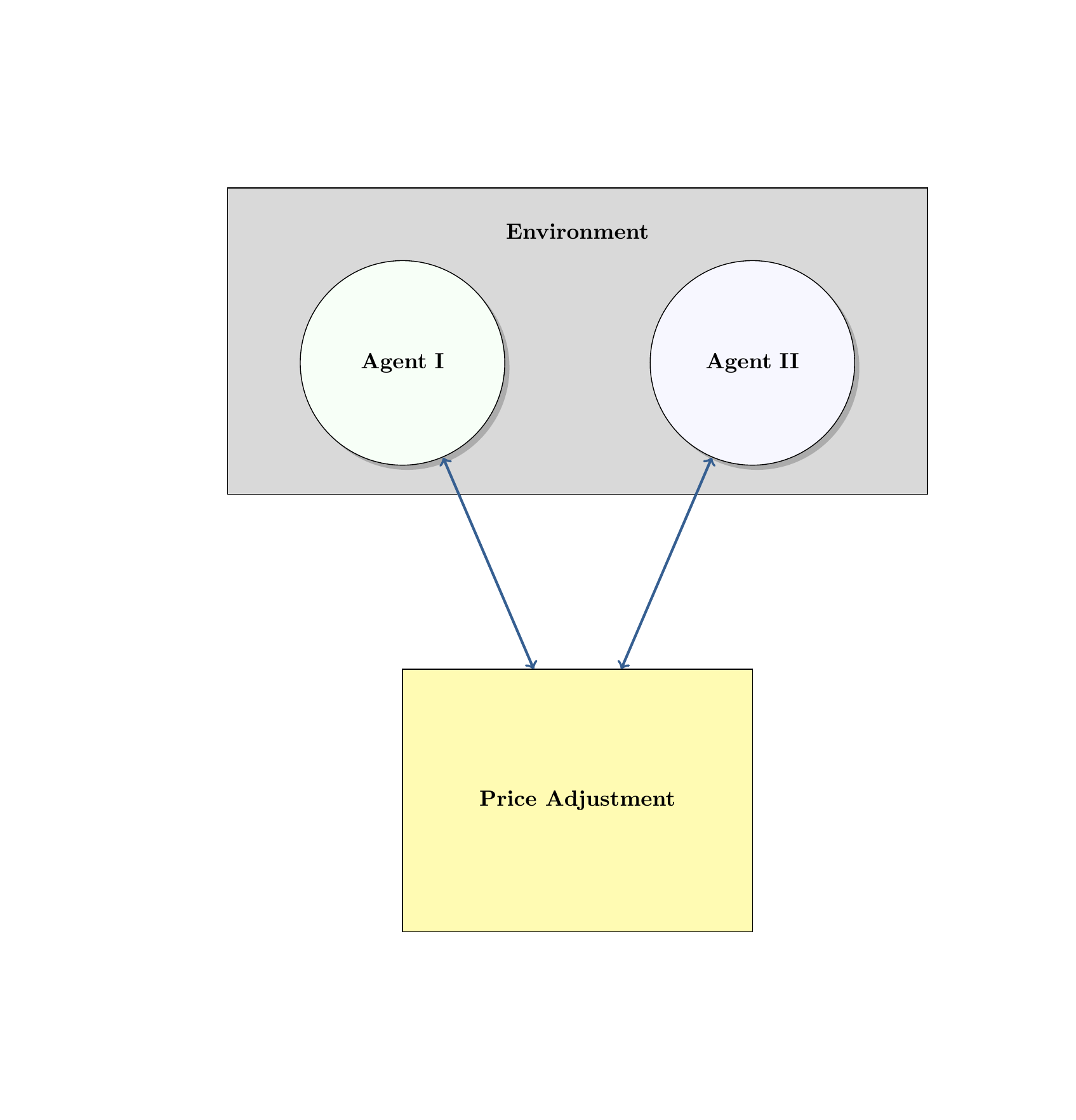}
\caption{Schematic picture of abstract ABEM model.}\label{framework}
\end{center}
\end{figure}

We review the recently introduced abstract agent-based economic market (ABEM) model \citep{trimborn2018sabcemm}.
The authors introduce a universal meta-model which helps to create, compare and categorize ABEM models.
The core idea is to define building blocks which are universal for most ABEM models.
These building blocks are agent design, market mechanism and environment. 
By agent design we mean the precise definition of an agent in each model.
The market mechanism can  be interpreted in the broadest sense as a rule which fixes a price of a good or stock at a financial market or between the agents.
The last building block is the environment, which can be seen as an additional coupling or spatial correlation among agents.
A schematic picture of this meta-model is given in Figure \ref{framework}.
In the following we will give specific examples of each building block.
We dispense on a rigorous mathematical definition of each building block as done in \citep{trimborn2018sabcemm}.

\paragraph{Price Adjustment}
We aim to specify what we mean by price adjustment process.
We define such a mechanism as a law which fixes a price of a certain good, stocks or bonds on a market.
Such a market may consist of all agents or an interaction of a subset of all agents.
Examples of a price adjustment process that considers all agents is a stock market which fixes the price for all agents.
An example of a price process that only considers a subset of agents may be a binary trade between agents or an auction.
Clearly, by this definition of a pricing process we implicitly define a financial agent as an actor on the market equipped with a personal supply or demand. 
Exemplified, we present a general disequilibrium market model implemented in many ABEM models \citep{trimborn2018sabcemm}.\\
We define the microscopic excess demand of all agents, given by demand minus supply.
These microscopic excess demands $ed_i, i=1,...,N$ of all agents can be aggregated to an aggregated excess demand $ED$.
$$
ED:=\frac{1}{N} \sum\limits_{i=1}^N ed_i.
$$
For a rigorous definition of aggregated excess demand we refer to \citep{mantel1974characterization, debreu1974excess, sonnenschein1972market}.
A general disequilibrium market model build on the idea of Beja and Goldman \citep{beja1980dynamic} is given by:
   \begin{align}\label{GenModApp}
 S_{k+1} = S_k+  F(S_k, ED_k)+ G(S_k,ED_k)\ \eta,
 \end{align}
Here, $\eta$ is a Gaussian distributed random variable and \eqref{GenModApp} is a difference equation. 
The index $k\in\N$ is the discretized time steps ($S_k=S(t+k\ \Delta t)$ for a fixed initial time $t$ and time step $\Delta t>0$).
Furthermore,  $F,G$ model arbitrary functions.
Many price adjustment processes of ABEM models are special cases of model \eqref{GenModApp}, for example the models presented in \citep{day1990bulls, alfarano2008time, lux1995herd, chiarella2002speculative, chiarella2005dynamic, chiarella2006asset, chiarella2007heterogeneous, challet2001stylized, zhou2007self, andersen2003game, harras2011grow, sornette2006importance, kaizoji2002dynamics, palmer1994artificial, bouchaud1998langevin, cont2000herd, cross2005threshold, cross2007stylized, cross2006mean, dieci2006market, farmer2002price, lux1999scaling, lux2000volatility, de2005heterogeneity}.

\paragraph{Agent Design}
The agent design differs for each field of applications.
Frequently used quantities which characterize the financial agent is wealth, investment propensity or agent's excess demand. 
In the following we present examples of agent's excess demand.
Many ABEMM models such as \citep{chiarella2006asset, beja1980dynamic, hommes2001financial, hommes2006heterogeneous, lux1995herd, franke2009validation} only consider two agents. 

\begin{example}\label{microdemand}
Frequently used financial agents in ABEM models are fundamentalist and chartists  \citep{lux1998socio, hommes2006heterogeneous}.
A fundamental agent believes that to every stock or good there is a fair market price (or fundamental value) $P^F>0$ and that the market price will converge to this fundamental value.
Hence, a fundamentalist wants to sell stocks if the stock price is above the fundamental value and he buys stocks if the market price is below the fundamental value.
Hence the agent's demand can be calculated as follows
$$
ed^F(t_k):= a\ (P^F(t_k)-P(t_k)),
$$
for a positive weight $a>0$ and a logarithmic stock price $P(t_k)\in \R$ with the discretized time  $t_k:= t_0+k\ \Delta t,\ \Delta t>0\ k\in\N$.
Here, $t_0>0$ denotes the initial time. A chartist assumes that the future stock return is best approximated by extrapolating past returns.
One simple example of such an investor is
$$
ed^C(t_k):= b\ (P(t_k)-P(t_{k-1})),
$$
for a positive weight $b>0$. 
Examples of such agents can be found in the models \citep{chiarella2006asset, beja1980dynamic, hommes2001financial, hommes2006heterogeneous, lux1995herd, franke2009validation}. 
\end{example}

In the previous example we presented two possible excess demands of two financial agents.
In the following we provide an example of how these excess demands can be aggregated into the aggregated excess demand.

\begin{example}
A simple example is the Franke-Westerhoff model \citep{franke2009validation, franke2011estimation, franke2012structural} where the population of agents is fully described by two agents ($N=2$), a chartist and a fundamentalist.
Hence, the excess demand is defined as
$$
ED^{FW}(t_k)= \frac12 \left( ed^{FW}_C(t_k)+ed^{FW}_F(t_k)\right). 
$$
In comparison to the general chartist and fundamental demand defined in Example \ref{microdemand} the Franke-Westerhoff models considers time dependent weights $a,b$ and additionally there is a random variable added. 
\end{example}

Then the price adjustment process as defined in the previous paragraph may utilize the aggregated excess demand in order to fix the price.
Clearly the above ideas have been generalized for $N$ agent designs \citep{cross2005threshold, harras2011grow, chen2012agent}. 
Especially we aim to emphasize that the previous example does not nearly cover the full range of possible agent designs.

\paragraph{Environment}
The previously introduced building blocks, agent design and price adjustment seem to be a natural structure in agent-based models.
We aim to introduce the notion of \textit{environment} which is to us the third building bock.
This concept  has been first introduced in \citep{trimborn2018sabcemm} and needs to be explained.
\\ \\
An environment subsumes any additional coupling, besides of the coupling via the price adjustment process, between the agents.
The most famous example possibly is herding, which is frequently used in ABEM models  \citep{alfarano2005estimation, kirman2001microeconomic, kirman1993ants, cross2005threshold, franke2012structural}.
Further examples are any spatial structure e.g. a network structure of agents or any prioritization of special financial agents \citep{ausloos2015spatial, alfarano2009network, alfarano2008should, harras2011grow, gurgone2018effects}.
As an example we aim to present the herding mechanism of the Cross model \citep{cross2005threshold}. 
\begin{example} We define the herding mechanism of the Cross model \citep{cross2005threshold}.
Each agent is described by a herding pressure $c_i>0$ and the evolution reads
\[ \begin{cases} c_i(t+\Delta t)= c_i(t)+ \Delta t |ED(t)|,& \text{if}\ \sigma_i(t)\ ED(t)<0\\
			c_i(t+\Delta t)=c_i(t),& \text{otherwise}.
			 \end{cases}
\]
Thus, the herding pressure is increased if the investment decision of the agent $\sigma_i\in\{-1,1\}$ has the opposite sign as the aggregated excess demand $ED$. 
This situation corresponds to the fact that the agents' position is in the minority. 
The agent switches position if the herding pressure $c_i$ has reached a threshold $\alpha_i>0$. After a switch the herding pressure gets reset to zero. 
This herding mechanism leads to additional coupling beneath the agents, introduced by the coupling with the aggregated excess demand. 
\end{example}
Finally, we aim to stress that these environments seem to be crucial in the generation of stylized facts. This seems to be natural since such an additional coupling often models behavioral aspects of agents e.g. herding. This coupling often leads to additional correlations among agents which may lead to clustering phenomena. 

%% file: conclusion.tex
\section{Conclusion}
\label{sec-conclusion}

In the first part we have given an overview of the most perceived stylized facts, namely fat-tails in asset returns, volatility clustering and absence of auto-correlation. 
Additionally, we have presented empirical results of DAX, Dow Jones and S\&P data.
We established that the excess kurtosis is a very volatile measure and heavily changes between different indices. 
Furthermore, we concluded that even the time horizon has an substantial impact on several statistical measures.
\\ \\
In the second part of the paper we have given an introduction to agent-based modeling.
After a short literature study we presented a short historical overview and reported major developments in this field of research.
Finally, we reviewed a recently introduced abstract ABEM model.
This model subdivides ABEM models in three building blocks.
Such an abstract formulation may help to create new models or compare existing ABEM models.
A detailed categorization of known ABEM models in this abstract framework is left open for future research. 